\documentclass[10pt,conference]{IEEEtran} 
\IEEEoverridecommandlockouts
\usepackage{cite}
\usepackage{amsmath,amssymb,amsfonts}
\usepackage{algorithm}
\usepackage{algorithmicx}
\usepackage{algpseudocode}
\algrenewcommand\algorithmicrequire{\textbf{Input:}}
\algrenewcommand\algorithmicensure{\textbf{Output:}}

\algnewcommand{\Input}{\Require}
\algnewcommand{\Output}{\Ensure}
\usepackage{graphicx}
\usepackage{subcaption}
\usepackage{textcomp}
\usepackage{xcolor}
\usepackage{makecell}
\usepackage{url}
\usepackage{listings}
\usepackage{graphicx}
\usepackage{booktabs}
\usepackage[normalem]{ulem}
\usepackage{tcolorbox} 
\usepackage{multirow}

\usepackage{xcolor} 
\usepackage{soul} 
\usepackage{amssymb} 
\usepackage{adjustbox} 
\usepackage[switch, mathlines]{lineno} %

\usepackage{listings}
\definecolor{dkgreen}{rgb}{0,0.6,0}
\lstdefinelanguage{diff}{
  language=Java,
  morecomment=[f][\color{blue}]{@@},     
  morecomment=[f][\color{red!60!black}]-,         
  morecomment=[f][\color{green!40!black}]+,       
  morecomment=[f][\color{magenta}]{---}, 
  morecomment=[f][\color{magenta}]{+++},
}

\usepackage{xspace} 
\newcommand{\toolname}{{\sc SemaDiff}\xspace}
\newcommand{\toolnamerand}{{\sc SemaDiff$_{Rand}$}\xspace}

\definecolor{dkgreen}{rgb}{0,0.6,0}
\definecolor{gray}{rgb}{0.5,0.5,0.5}
\definecolor{mauve}{rgb}{0.58,0,0.82}
\newcommand{\chan}[1]{{\color{black} #1}}

\newboolean{showcomments}
\setboolean{showcomments}{false} 
\ifthenelse{\boolean{showcomments}}
 { \newcommand{\mynote}[2]{
      \fbox{\bfseries\sffamily\scriptsize#1}
        {\small$\blacktriangleright${{{#2}\bf }}$\blacktriangleleft$}}}
        { \newcommand{\mynote}[2]{}}

\newcommand{\ak}[1]{\mynote{Ahmed}{\hl{#1}}}

\def\BibTeX{{\rm B\kern-.05em{\sc i\kern-.025em b}\kern-.08em
    T\kern-.1667em\lower.7ex\hbox{E}\kern-.125emX}}
\begin{document}

\title{\textit{SemaDiff:} Identifying Semantic-Changing Commits with Generated Code and Tests
}

\author{

\IEEEauthorblockN{
Maha Ayub$^1$, 
Michael Konstantinou$^1$,
Ahmed Khanfir$^{2,1}$,
Nikolaos Tsantalis$^3$, 
Mike Papadakis$^1$
}

\IEEEauthorblockA{
\{maha.ayub, michael.konstantinou, michail.papadakis\}@uni.lu \\
ahmed.khanfir@ensi-uma.tn \\
nikolaos.tsantalis@concordia.ca
}

\IEEEauthorblockA{
$^1$ \textit{SnT, University of Luxembourg}, Luxembourg \\
$^2$ \textit{RIADI, ENSI, University of Manouba}, Tunisia \\
$^3$ \textit{Concordia University}, Canada
}

}


\maketitle

\begin{abstract}
Distinguishing semantic-preserving commits from changing ones remains an open challenge in software repository mining. While existing approaches detect refactoring commits accurately, they cannot ensure that a commit is purely semantic-preserving, without any interleaving behaviour-changing modification. 
This limitation can impact several tasks, such as debugging, fault localisation, bug dataset construction, rollback analysis, and bug fixes backporting. 
To fill this gap, we propose \toolname, a novel approach for identifying semantic-preserving commits through behaviour-based analysis; comparison of similar test execution on pre- and post-commit versions. 
As code impacted by the refactoring is often hard to test and different accross both versions, we propose generating additional calling methods to that code, which serve as testing target.
Given a commit, \toolname analyses the diff to identify modified code and extracts unchanged dependent code that calls it. 
It then generates an additional dependent class using a large language model to exercise the changed code in both versions, and automatically generates tests for the dependent code. This way, we obtain the same tests for the different code versions, enabling the behavioural-difference detection.
The commit is classified as semantic-preserving only if all generated tests produce identical outcomes across the two versions. 
To evaluate \toolname, we construct and annotate manually a dataset of \chan{183} 
commits, gathered from well-known open-source Java projects. 
The obtained results show that \toolname distinguishes accurately semantic-preserving from -changing commits in \chan{$\approx$76\%} of the cases, with a \chan{100\%} precision in semantic-changing commit detection. 

 
\end{abstract}



\section{Introduction}
\ak{add/update info about dataset + baselines and comparison + focus on the added value in the identification of semantic changing, rather than accuracy...}
Software 
is usually not a one-time artifact, but is continuously evolving through the addition of new features, the fixing of bugs, and the optimization of existing code, etc. 
 These form typical activities of developers, committing and pushing daily code modifications to their projects, which are usually hosted centrally, i.e. in Version Control Systems like Git~\cite{Github}. 
The history of code changes is kept carefully, as it can be useful in achieving several software engineering tasks. 
For instance, to apply a fix on a previous version (also known as backporting~\cite{backporting}), to identify the latest stable version, or to identify bugs' root causes, practitioners need to go through the history of commits, determining which one caused the target change~\cite{10.1145/3576915.3623188}.
To simplify this search task, developers tend to include hints related to the commit type in the commit messages, e.g. \texttt{bug-fix}, \texttt{refactoring}, or \texttt{feature}, etc.
Although helpful, this type of labeling is not always provided, and often not reliable, as developers may mention only the main intended changes in the commits and not others~\cite{10.1145/3266237.3266260}.
This is because developers often introduce several changes (of different types) in one commit, making commit-message level search impractical.
In fact, a recent study on the impact of refactoring on code evolution \cite{refactoringOcuurence} indicates that approximately 74\% of the commits involve refactoring. 

Moreover, the inherent complexity of some refactoring makes them more error-prone and difficult to control\cite{10.1145/1287624.1287651}\cite{7962330}. For instance, the report \cite{10.1007/978-3-642-39038-8_23} shows that nearly 80\% of the changes that break client applications are API-level refactoring edits. In addition, 77\% of the respondents in the survey of Microsoft developers~\cite{10.1145/2393596.2393655} indicate that refactoring frequently introduces subtle bugs and functionality regressions. Hence, the need to validate whether code refactorings 
preserve the behaviour or not.
For instance, Figure~\ref{motivation-example} illustrates a refactoring commit, where the variable is inlined directly into the return statement as \texttt{Array.asList(Tokens)}. 
However, these changes make \texttt{getTokenList()} no longer return an independent, mutable copy, but instead return a fixed-size view that is backed by the original array. As a result, it can cause runtime exceptions (\texttt{UnsupportedOperationException} on add or remove) and introduce aliasing between the returned list and the internal tokens array. The issue is also reported on the Apache project's issue tracking system~\cite{ASFjira}.

These reasons motivated much research to focus on detecting refactoring commits. 
The detection of such commits operates typically by static-analysis and comparison of pre-and post-versions of commits in order to identify refactoring operators. To this end, refactoring operators have been proposed and tuned through several studies, leading to stable and mature pattern-based approaches, like RefactoringMiner\cite{Refminer} and Refdiff\cite{Refdiff}, etc., which work on the static level. 
Although very efficient in detecting refactorings, together with their type and location, these approaches do not give any information about whether these changes preserve the semantics of the code or not. 
In fact, they detect whether a commit introduces refactorings, i.e. matching one or many refactoring types, but they do not give any information on whether other than structural changes have been introduced.
Therefore, they are not able to ensure whether these refactorings are indeed semantic preserving or not, e.g., buggy or bug-prone.
Aiming at distinguishing purely refactoring commits from others, RefactoringMiner authors proposed --Purity Checker~\cite{purity-checker}-- which checks whether all commit changes are categorized as refactoring or not. 
Although lightweight and performant, it remains of limited scope as it handles only 9 refactoring types and does not investigate behavioral changes, but operates solely on the syntax of the code.
In addition to these tools, developers can rely on 
unit tests, as well as regression test generation tools 
to detect 
behavioural changes. 
In fact, obtaining different execution results of same test-suite between both versions (pre- and post-commit) indicates a semantic change in the program~\cite{ojdanic2023syntactic}.
However, these existing or auto-generated tests would often not compile or execute on both versions 
for multiple refactorings, e.g. tests that call the method \texttt{getTokenList} would not compile if the refactoring renames it to \texttt{getTokenLst}. 
Even when ignoring these renaming cases, a recent study~\cite{faultregressiontools} showed that automated test generators perform relatively poorly in this task, uncovering less than \chan{50\%} of the studied refactoring faults.

\begin{figure}
    \centering    
\begin{lstlisting}[language=diff, captionpos=b ]
public List<String> getTokenList() {
    checkTokenized
-   final List<String> list = new ArrayList<>(tokens.length);
-   Collections.addAll(list, tokens);
-
-   return list;
+   return Arrays.asList(tokens);
}
\end{lstlisting}
\caption{Example of semantic changing refactoring commit. \ak{we can keep its if we have space}}
\label{motivation-example}
\vspace{-1em}
\end{figure}




In this work, we propose \toolname, a novel approach that identifies semantic changes by comparing the execution results of the same tests on both commit versions.
To address the issues of test compilation and execution on different codes, \toolname targets the testing on the unchanged code by the commit, which calls the modified one.  
To enable and improve testing coverage even when modified code is not called, the approach generates additional dependent classes. Finally, it generates test cases targeting this dependent code, and thus, exercising the refactored one. 
To evaluate the effectiveness of our technique, we collected and annotated \chan{183} recent refactoring commits from seven different repositories. 
Our results show that \toolname 
determines whether refactoring commits preserve semantics or not in \chan{$\approx$76\%} of the cases, with a precision or \chan{100\%}. 

In this paper, we make the following main contributions:
\begin{itemize}
    \item We propose \toolname, a novel approach for identifying Semantic-changing commits.
    \item We construct and propose a diverse dataset\footnote{The replication package and dataset will be made publicly available upon acceptance.} of \chan{183} commits containing both semantically preserving and non–semantically preserving changes, all manually inspected and labeled.
    \item We provide empirical evidence on the effectiveness of \toolname in distinguishing semantic-changing and -preserving commits. 
    \item We provide a comprehensive empirical evaluation study of \toolname, validating its design components and discussing its limitations. We study and discuss the effectiveness of static-analysis baselines such as, large language models and PurityChecker, in this same task.
\end{itemize}

\section{Background and related work}
\label{sec:background}



\subsection{Refactoring detection}
\begin{figure*}[h]
\vspace{-0.5em}
  \centering
  \includegraphics[width=0.95\textwidth, trim=10cm 10cm 10cm 10cm, clip]{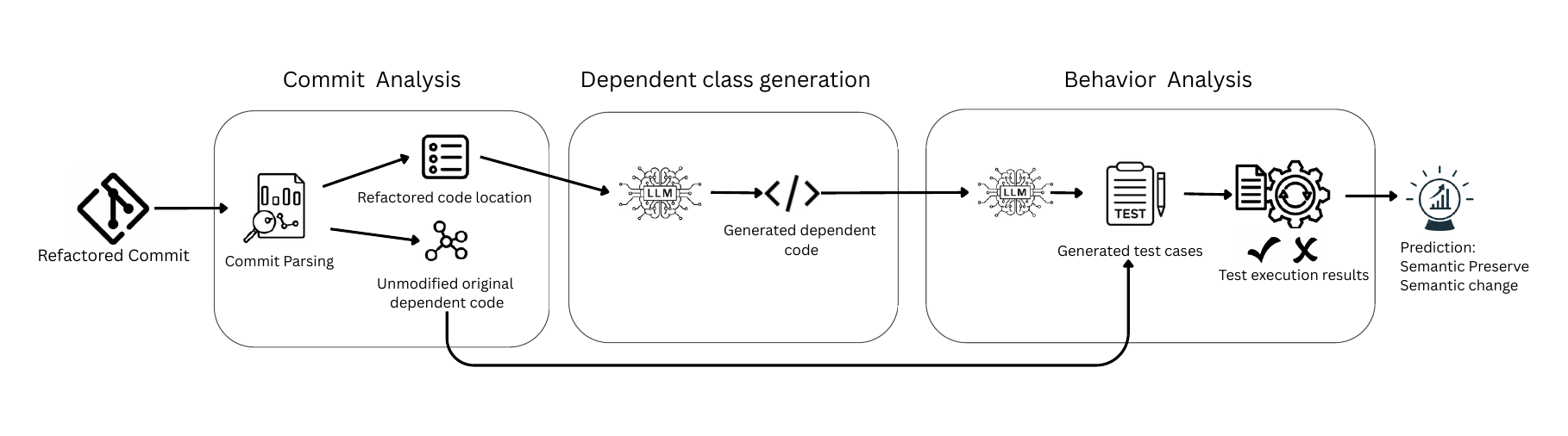}
  \vspace{-2.0em}
  \caption{Refactoring semantic detection flow.}
  \label{Main-flow}
  \vspace{-1.0em}
\end{figure*}
Refactoring is the process of restructuring existing source code to improve or adapt its 
design and structure while preserving its 
behavior and semantics \cite{Ops_2022}. However, in practice, refactoring may carry risks and can lead to bugs \cite{6392107, 10.1145/3368089.3409695}, as refactoring edits are in many cases tangled with edits related to other maintenance tasks \cite{dias2015untanglingfinegrainedcodechanges, 10.1145/3540250.3549171}. Moreover, refactorings are often carried out manually, even though automated refactoring tools exist. One study \cite{10.1007/978-3-642-39038-8_23} shows that developers perform manual refactorings more frequently than automated ones, and around 90\% of these manual refactoring tasks are prone to errors \cite{10.1145/2568225.2568280}. Furthermore, automated refactoring tools themselves are not entirely reliable in guaranteeing semantic preservation. For instance, one study \cite{10.1145/1287624.1287651} found that testing refactoring functionalities on generated Java programs revealed more than 20 previously unknown bugs in the refactoring implementations of the Eclipse and NetBeans IDEs. Several studies \cite{SafeRefactor}\cite{10.1145/3747289} have addressed issues related to refactoring correctness; however, their primary focus has been on identifying bugs introduced by refactoring engines rather than on comprehensively verifying semantics preservation after refactoring.

In the context of software repository mining, refactoring commits are those that contain such structural modifications. 
These refactoring commits are typically detected through static analysis of the differences between the pre- and post-commit versions, and the identification of predefined refactoring-patterns~\cite{10.1145/3705309}. 
A long history of research in this field led to mature rule-based detection approaches and reliable tools, e.g., RefactoringMiner \cite{9136878} and RefDiff \cite{7962377}.
Although they can accurately detect refactoring activities and changes within a commit, they do not assess whether additional modifications (such as bug fixes or new feature implementations) have been made, and therefore cannot determine if the refactoring commit is semantically preserving or not.



\subsection{Refactoring semantic verification }
To verify semantic preservation after refactoring, various regression testing tools such as Randoop and EvoSuite are commonly used to automatically generate test suites and validate program behavior. However, these tools have notable limitations. A study \cite{10.1145/3559744.3559745} demonstrates that automatically generated regression suites fail to detect all faults introduced by Extract Method refactorings. Therefore, relying solely on such regression tools to ensure semantic equivalence after refactoring provides limited assurance.

Beyond refactoring engines and regression testing tools, Large Language Models (LLMs) have recently been adopted in software engineering tasks. LLMs can generate functional code directly from natural language problem descriptions with relatively high success rates \cite{10.1145/3691620.3695527}. They are also used to generate automated unit tests \cite{10.1145/3663529.3663801}. However, depending exclusively on manually written or LLM-generated unit tests remains insufficient. Refactoring may alter the internal structure of a program without changing its observable behavior, potentially breaking the alignment between source code and its associated unit tests. Since unit tests often depend on specific structural elements that refactorings modify, they can become outdated or invalid even when program semantics are preserved \cite{10.1145/2972206.2972223}. In a similar context, some tools employ fuzzing techniques to validate refactoring. For example, one approach \cite{dristi2026differentialfuzzingbasedevaluationfunctional} uses fuzzing to automatically generate test cases that assess LLM-based refactoring. However, under the same unit test conditions, this method does not work effectively for rename refactoring, and the authors primarily concentrate on validating LLM-generated refactorings rather than including all refactorings produced manually or by other tools. 

In addition to regression testing, fuzzing and unit testing approaches, ChangeGuard \cite{changeguard} proposes a technique for preserving semantics in refactored code. It applies the learning-guided execution technique that runs the original and refactored versions in isolation on the same inputs. This technique is effective for relatively simple programs involving basic data types and deterministic behavior, but such assumptions rarely hold in real-world software systems. In contrast, \toolname operates directly on real-world refactoring commits, supports diverse refactoring types, and is capable of handling complex codebases to perform robust semantic verification.



\section{Approach}
\label{sec:approach}
\toolname operates in three steps as illustrated in Figure ~\ref{Main-flow}. 
For a given commit, 
\begin{enumerate}
    \item \textbf{Commit analysis:} \toolname parses the commit-changes, in order to identify modified locations and their corresponding unmodified calling code.
    \item \textbf{Dependent code generation:} it generates code, i.e., a class, that calls the modified portions of the code and runs similarly on both commit versions.
    \item \textbf{Behavioral analysis:} generates tests targeting code that calls the program-modified parts, then runs them on both versions, and compares their execution results. The commit is said to be \textit{semantic-preserving} if we obtain the same results, and \textit{semantic-changing}, otherwise. 
\end{enumerate}

\subsection{Commit Analysis}
As it is often difficult, even impossible, to run the same tests on different codes, we target the testing on the unchanged portions of the code, which exercises the changes.
To do so, we start by extracting the list of modified source-code files from the input commit.
If the list is not empty, \toolname checks out both pre- and post-versions of the commit and 
\ak{add in the dataset description that we focus on Java files and keep only commits having code modifications}
analyses their abstract syntax tree, in order to construct the call graphs of the modified code by the commit.
This way, we obtain the list of the calling dependent classes on both versions. 
Finally, we filter out any modified dependent code, keeping only the unchanged portions between both commit versions. 



\begin{figure}
    \centering
     \begin{subfigure}[b]{0.49\textwidth}
     \centering
\begin{lstlisting}[language=diff]
-  private static <T> T[] defensiveCopy(final T[] src) {
+  private static <T> T[] clone(final T[] src) {
        return src != null ? src.clone() : null;
    }
  public Configuration[] getConfigurations() {
+  return defensiveCopy(configurations);
-  return clone(configurations);
    }
\end{lstlisting}
\vspace{-1.0em}
\caption{Example of method renaming refactoring commit changes.}
\vspace{0.6em}
\label{subfig:dependent-rename-refactoring}
\end{subfigure}

     \begin{subfigure}[b]{0.49\textwidth}
     \centering
\begin{lstlisting}[firstnumber=1]
public class ConfigurationPropertiesFactoryBeanDependent {

 private final ConfigurationPropertiesFactoryBean bean = new ConfigurationPropertiesFactoryBean();

 public Configuration firstConfigAfterMutatingInputArray(final Configuration originalFirst, final Configuration replacement) {
        final Configuration[] input = new Configuration[] { originalFirst };
        bean.setConfigurations(input);
        // Mutate the original array after setting into the bean
        input[0] = replacement;
        final Configuration[] current = bean.getConfigurations();
        return current != null && current.length > 0 ? current[0] : null;
 }
}
\end{lstlisting}
\vspace{-0.7em}
\caption{Generated dependent code by the model calling.}
\label{subfig:dependent-class-example}
\end{subfigure}
\hfill
\vspace{-0.6em}
    \caption{Example of generated dependent. }
    \vspace{-0.6em}
\end{figure}





\subsection{Dependent class generation}
\label{subsec:dependent_class_gen}

As the modified code is often rarely called, or not at all, the previous step fails at providing sufficient or any Unchanged dependent code to test.
This is often the case for libraries and frameworks having uncalled code, which is meant to be used by other applications. 
To alleviate this shortage and enable semantic-change analysis in such cases, we propose the artificial code generation of dependent calling code.
To achieve this, we feed a generative language model, i.e. ChatGPT, with a prompt, the commit diff and the modified class name, in order to generate a dependent class that compiles and calls the modified code on both versions. 
To ensure that the code compiles, we instruct the model to use only \texttt{public} methods and include all necessary imports.
These instructions are given to the model as the following prompt:
 \begin{tcolorbox}[colback=gray!2, colframe=green!50,boxrule=0.4pt,arc=1pt,left=6pt,right=6pt,top=4pt,bottom=4pt,before skip=8pt,after skip=8pt]
\textit{"Given a commit diff, generate Java source files:
         A dependent class that uses public APIs and compiles on both parent and modified versions.
                Core Rules:
                - Output MUST compile successfully against both the parent and modified versions.
                - Use Java X syntax only
                - The file MUST start with the correct package declaration and required imports.
                - Prefer stable public APIs over internal/private APIs."}
\end{tcolorbox}
Once the code is generated, we add it to both versions of the project and compile it. If it does not compile, we repeat code generation, i.e. by asking the model again; otherwise, it is kept as additional target-dependent code to test.


Subfigure~\ref{subfig:dependent-rename-refactoring} illustrates an example commit diff\footnote{\url{https://github.com/apache/commons-configuration/commit/972b998291b26b5acf011c78e6c9fa3fbaa6f397}}, which renames the \texttt{private} method \texttt{defensiveCopy} to \texttt{clone}. 
As can be seen in~Subfigure~\ref{subfig:dependent-class-example}, the corresponding generated class executes it indirectly by calling the \texttt{public} method \texttt{getConfigurations}, instead of calling it directly. In addition, the generated method \texttt{firstConfigAfterMutatingInputArray} takes two parameters, exercising the target method with different values. 



\subsection{Behavioral Analysis}

This step is key to the main goal of our approach, which is detecting semantic changes in commits. 
It focuses on generating and running test cases on the dependent classes, targeting edge cases in the modified code. 
To do so, we provide original (developer-written) and generated dependent code -- obtained in the previous steps -- together with the commit diff and ask a generative model, i.e. ChatGPT, to generate tests that would execute differently on the two versions if the commit is semantic-changing and similarly otherwise. 
This way, commits that include only pure refactoring transformations, such as renaming 
or signature updates, etc, the dependent usage remains semantically equivalent, and the same test suite is expected to execute similarly on both versions. 
Moreover, we believe that passing the diff to the model would better guide it in focusing the testing on the modified code, and thus, covering more thoroughly related corner cases.
These inputs are passed to the model with the following prompt

 \begin{tcolorbox}[colback=gray!2, colframe=green!50,boxrule=0.4pt,arc=1pt,left=6pt,right=6pt,top=4pt,bottom=4pt,before skip=8pt,after skip=8pt]
\textit{"Be conservative: if the diff may change observable behavior in any class, generate tests that expose whether functionality is preserved.
                - For pure refactorings, renames, formatting changes, or API-equivalent changes, generate passing tests only.
                - For added, removed, or changed functionality, generate tests whose assertions fail when behavior is not preserved.
                - Return only the most confident test cases supported by the diff.
                - If behavior changed, assert the visible behavior directly.
                - If the change is pure refactoring, generate only passing tests."}
\end{tcolorbox}

After generation, the test suite is added to the project and compiled on both commit versions.
If the compilation fails for any of the versions, we send the test cases with the error stack-trace and ask it solve it. 
We repeat this compilation validation feedback loop with the model until we obtain a compiling test-suite and up to a maximum of three times.
Finally, we execute the tests on both versions and compare their results; a semantic change is detected if any of the tests execute differently in one of the versions. Otherwise, we consider that the behaviour remained unchanged by the commit.

\section{Research Questions}
\label{sec:rqs}
As we aim at distinguishing between semantic-preserving and semantic-changing  commits, we start by asking:
    \begin{description}
    \item[RQ1] \emph{(Efficiency)}: How efficient is \toolname{} in distinguishing semantic-preserving from semantic-changing commits?
    \end{description}
    
    We run our approach on a set of refactoring commits and evaluate its capability in detecting whether they verify the semantics preservation or not. 
    
    To assess the contribution of the generative component in \toolname effectiveness, we ask: 

    \begin{description}
    \item [RQ2] \emph{(Generative Language Model)}: How does \toolname perform when using different LLMs? 
    \end{description}

    We create similar variants of \toolname using other LLMs than ChatGPT and compare their efficiency in distinguishing semantic-preserving commits from changing ones. 
    The answer to this question provides quantitative results on the tool's performance variation under different code generation models.

    Next, we check the importance of    \chan{changes-dependent code generation (more details in Subsection~\ref{subsec:dependent_class_gen})}, and ask:
    
    \begin{description}
    \item [RQ3] \emph{(Dependent class generation)}: What is the contribution of Generated Dependent Class in \toolname's efficiency? 
    \end{description}
    We compare the efficiency of \toolname generated dependent class with the real dependent class available in the repository.
    The aim is to validate the approach-design decision, bringing empirical evidence that, the use of a generated dependent class in our approach enables its high detection capabilities.
    
 \toolname relies on test execution results to detect semantic changes. We thus investigate its effectiveness variation when using tests generated by different approaches. Hence we ask:
    \begin{description}
    \item [RQ4] \emph{(Tests generation)}: What is the impact of the choice of automated test suite generation on identifying commit behavioral changes?
    \end{description}
    
    We compare the efficiency of \toolname with a baseline-approach that employs a rule-based automated-test generation technique, i.e., Randoop, instead of generative language models. 
    To have a fair base of comparison, both approaches are run on the same subjects, targeting the same code and with the same effort-budget, i.e., in terms of generated tests. 
    These results validate our approach's choice of generating tests by large language models, yielding test suites that better detect program semantics, and thus, higher semantic-changes detection capabilities. 
    To explain the higher performance of our approach, we extend our study by analysing tests from both approaches, and particularly the semantic changes, revealing tests from others, in terms of code coverage. 
    This allows having a better understanding of the impacts and differences between the test generation approaches. 
Thus, we ask:
    \begin{description}
     \item[RQ5] \emph{(Comparison with static analysis approaches)} How efficient are 
     PurityChecker and LLMs in detecting semantic-changing commits? How do they compare with \toolname?
    \end{description}
    To answer this question, we compare the effectiveness of four static-analysis baselines in distinguishing between semantic-preserving and -changing commits, which operate without performing any dynamic behavioral analysis, i.e. no code and test generation. 
    This comparison enables a better assessment and positioning of \toolname, endorsing the importance of dynamic-analysis in detecting corner-case semantic-changes.

\section{Experimental Setup}
\label{sec:setup}

\begin{table*}[htbp]
\vspace{-0.5em}
\centering
\caption{Metrics of manually collected commits}
\vspace{-0.5em}
\label{tab:semantic_analysis}
\begin{tabular}{lcccccl}
\toprule
\textbf{Repository} & 
\textbf{Commits} & 
\makecell{\textbf{Semantic Preserving}} & 
\makecell{\textbf{Semantic Changing}} & 
\makecell{\textbf{Modified Java Files}} & 
\makecell{\textbf{Refactored Lines}} & 
\textbf{Time Span} \\
\midrule
commons-cli        & 29 & 12 & 17 & 35 &  1059 & 2023--2025 \\
commons-configuration    & 25 & 15 & 10 & 54 & 1728 & 2022--2025 \\
commons-io           & 13 &  4 & 9 & 17 & 764 & 2024--2025 \\
commons-lang       & 41 & 26 &  15 & 47 &  1754 & 2024--2026 \\
commons-net        & 21 & 19 & 2 &  29 & 1198 & 2024--2025 \\
chesslib    & 18 & 5 &  13 & 25 & 1613 & 2020--2026 \\
jsoup             & 36 &  7 &  29 &  55 &  2283 & 2024--2026 \\
\midrule
\textbf{Total}           & \textbf{183} & \textbf{88} & \textbf{95} & \textbf{262} & \textbf{10,399} & -- \\
\bottomrule
\end{tabular}
\end{table*}

\begin{figure*}[t]
    \centering
    \vspace{-0.6em}
    \begin{subfigure}[t]{\columnwidth}
        \centering
        \includegraphics[width=0.9\linewidth]{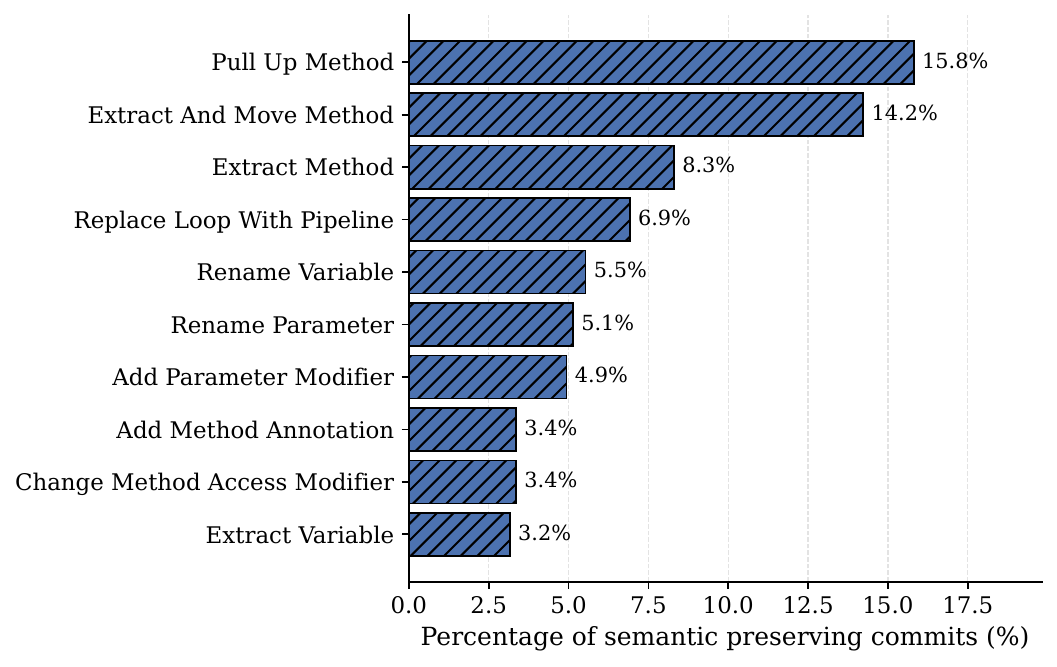}
        \caption{Semantically preserving commits}
        \label{refactoring-semantic-preserve}
    \end{subfigure}
    \hfill
    \begin{subfigure}[t]{\columnwidth}
        \centering
        \includegraphics[width=0.9\linewidth]{Figures/icse_refactoring_types_semantic_preserving_top10.pdf}
        \caption{Non–semantically preserving commits}
        \label{refactoring-non-semantic-preserve}
    \end{subfigure}
    \caption{Percentage of the top 10 refactoring types in semantically and non-semantically preserving commits}
    \vspace{-0.5em}
    \label{refactoring-types}
\end{figure*}

\subsection{Dataset \& Benchmark}
\label{subsec:dataset}
To create our commits dataset, we started by systematically fetching recent commits from \chan{7} open-source repositories.
As our approach heavily relies on Large Language Models in code and test generation, we believe that having recent commits can reduce the risk of data-leakage and seen-data during the model's training.
We acknowledge, however, that with the current multi-agent architecture of LLMs, and particularly their Retrieval-Augmented Generation (RAG)~\cite{lewis2020retrieval} components, they can access online resources at prediction time (post-training).
Nevertheless, from our RQ5 results in Subsection~\ref{RQ5}, we can see that LLMs are challenged by our dataset, not detecting all commits accurately.

The next systematic step consists of filtering not refactoring commits. While this step is intuitive for selecting semantic-preserving commits, it is not for the semantic-changing ones. 
In fact, to construct a solid and challenging set of these latters, we opted for tangled commits~\cite{herzig2011untangling} -- introducing not only behavioural modifications but also refactoring ones.
As a first quick filtering step, we systematically exclude non-refactoring commits based on commit message analysis, e.g., excluding those containing “bug fix,” or similar terms in the commit message.
As some commits may be invalid, e.g. requiring unavailable libraries, we checkout and compile all commits and discard those not compiling, ensuring the usability of the remaining set in our study.
Finally, we feed the obtained commits to \chan{RefactoringMiner~\cite{Refminer}} and include those in which at least one refactoring change is detected, obtaining about \chan{183} compiling commits. 

To obtain the ground truth, we annotate the commits, labelling each as either semantics-preserving or not. 
This is achieved through manual inspection of all dataset subjects by three authors (junior and senior engineers and researchers in software engineering) individually and independently. 
Every author is tasked to provide 1) a label (semantic-changing or not), 2) a small description of the commit intent, and 3) a confidence score of the labeling, between 0 and 5. Even when not confident, the authors were asked to attribute a label, i.e. the most likely, to every commit.

Once done, the authors met to debate 41 ambiguous cases and label them collectively. 
While many were easy to agree on, some required extensive analysis and check, i.e. test writing, presenting several implementation changes with very hard to spot behavioural changes or preservation.
Many of the discussed commits (15) were adding functionality without impacting the business logic of the program. 
For instance, making a \texttt{private} method \texttt{public} does not impact the behaviour of the program but opens an additional access to the methods functioning in that class. \ak{we could add an example here}
As we cannot ensure that this change does not impact the program functioning, particularly in the case of libraries, e.g. allowing for a misuse or a buggy scenario, we annotate any functionality addition as semantically-changing. 
Overall, the high number of commits further highlights the complexity of the task and thus, motivating further the goal of our work.
\ak{I am putting this in red because i am not sure if we need to mention it...}

\begin{figure}[t]
\vspace{-0.5em}
    \centering
  
  \adjincludegraphics[width=\linewidth, trim={{.00\width} {.00\width} {0.00\width} {.00\width}} ,clip]{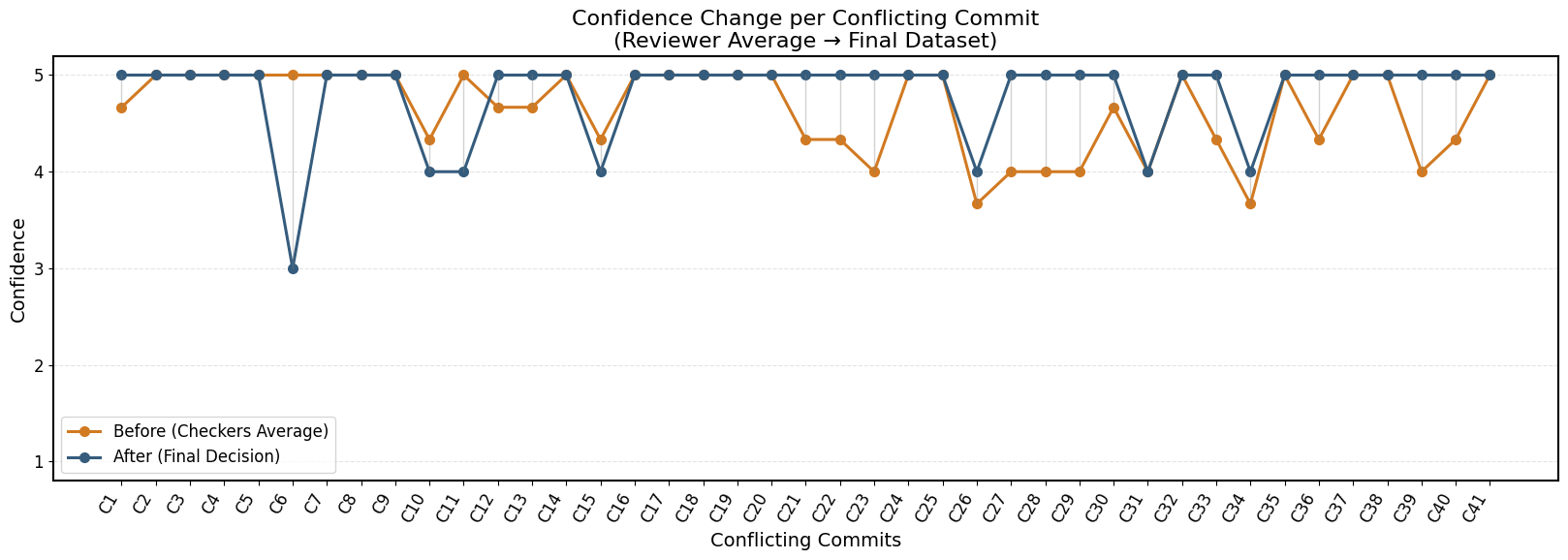}
    \caption{Confidence of manually labeling conflicting commits}
    \vspace{-0.5em}
    \label{manual-label-confidence}
\end{figure}

\ak{Maha: i don't think the confidence scores figure is adding much,... but i will keep it for now...}
Figure~\ref{manual-label-confidence} presents the average confidence scores assigned by the reviewers before and after discussion for the discussed commits.
When annotating individually and before any debate, the authors match the final labels in the majority of the cases, precisely in \chan{91.8, 88.5 and 88.5\%} of the commits. 

Our resulting labeled-dataset counts \chan{95} semantic-preserving and \chan{88} semantic-changing commits, introducing various types of refactoring as illustrated in
Figure \ref{refactoring-types}. The table \ref{tab:semantic_analysis} presents the information 
about the included commits in our dataset; repositories, the count of refactored files, applied refactorings, as well as the time span of the repositories.
\ak{Maha: we can keep these details if we still have space in the end.}

\subsection{Experimental Procedure}
\label{subsec:exp_proc}
To evaluate \toolname semantic-changes detection, we run it on our dataset and check its accuracy, precision and recall in distinguishing semantic-preserving and semantic-changing commits, which we compute as follows (\textit{answer to RQ1}):
\begin{equation}
Accuracy = \frac{TP + TN}{Total\;Commits}\times 100
\end{equation}
\begin{equation}
Precision = \frac{TP}{TP + FP}\times 100
\end{equation}
\begin{equation}
Recall = \frac{TP}{TP + FN}\times 100
\end{equation}

, where the semantic-changing commits constitute the positive subjects to identify. Consequently, $TP$, $TN$, $FP$ and $FN$ are defined as follows:
\begin{itemize}
    \item[TP:] refers to the number of \textit{True Positives} consisting of correctly predicted semantic-changing commits.
    \item[TN:] refers to the number of \textit{True Negatives} consisting of correctly predicted semantic-preserving commits.
    \item[FP:] refers to the number of \textit{False Positives} consisting of mispredicted semantic-preserving commits, as semantic-changing ones.
    \item[FN:] refers the number of \textit{False Negatives} consisting of not detected (missed) semantic-changing commits, and classified as preserving.
\end{itemize}
We compute also the execution failure rate as follows;
\begin{equation}
ExecFailure =
\frac{Failures}
     {Total\;Commits}
\times 100
\end{equation}, where $Failures$ refers to the number of commits for which the approach failed to execute or provide any result and the ${Total\;Commits}$ the number of commits in our dataset.
To account for these execution failures, we divide the number of accurate predictions by the ${Total\;Commits}$, including commits in which no prediction was given.

To evaluate the impact of the LLM selection (\textit{answer to RQ2}), we create three identical variants of it, using each a different model, namely, ChatGPT~\cite{chatgptpaper} (default, used to answer all research questions), CodeLlaMa~\cite{codellamapaper} and DeepSeek~\cite{deepseekai2024}. 
We run and compare their performance, using the same metrics and on the same commits from our dataset.
\ak{Maha: If not already done in the related works, we should add a 1 line description of every model and a reference.}
We also keep track of every prediction with the issuing model-variant and corresponding commit, to study eventual complementarity between the three models.
They are broadly adopted by researchers and practitioners, and have been proven efficient in coding and testing related tasks~\cite{LI2025103942,konstantinou2026well},~\ak{cite somenthing here} forming therefore good generative language model candidates for \toolname.

To assess the contribution and importance of the dependent-code-generation component of \toolname, we compare the semantic-changes detection capabilities of the tests generated when targeting generated code Vs those of the original code (\textit{answer to RQ3}). To enable this, we map every test execution result with its targeted class.
We also focus on semantic-changing commits, as it is often harder to write tests exposing such changes than passing on semantic-preserving ones.
More precisely, we compare the effectiveness of every approach in producing semantic-change-revealing tests for our commits.

To evaluate \toolname test generation component, we compare its contribution in detecting semantic changes with that of a rule-based \toolname version, which we note \toolnamerand (\textit{answer to RQ4}). 
This baseline relies on a well-established feedback-guided automatic test generator -- Randoop -- which, for a given Java code, generates a test suite capturing its behaviour, and enabling regression testing.
To ensure a fair base of comparison, we compare \toolname and \toolnamerand on the same original and generated target code.
To account for the stochastic nature of Randoop, we run it five times.
To investigate the performance between both test generation approaches, we compare test suites coverage of modified code by the subject commits, on the class- and line-level of granularity. 
\ak{i am not sure about the RQ... waiting for results...}

Finally, we compare the prediction performance of 4 baselines with the same setup; PurityChecker~\cite{purity-checker} and three LLMs, namely,  ChatGPT, CodeLlaMa and DeepSeek. 
PurityChecker, is a pattern-based extension to RefactorMinor, which takes as input the commit URL and checks whether the behavior is the same or not.
The three LLMs receive a prompt asking them to predict whether the commit preserves the semantics or not (without any code or test generation steps), given the same input as \toolname; the commit-url and diff. 
We do not report $ExecFailure$ rates for these approaches, as they all run successfully on our dataset. Instead, we report $NoResult$ rates for commits falling outside their scopes, i.e. PurityChecker covers a limited number of refactoring types and does not give any results for commits having unhandled ones.





\subsection{Implementation}
\label{subsec:Implementation}

We implemented \toolname in Java. We used 
Spoon~\cite{PawlakMPNS16} to construct the dependency graph between the modified (called) and the dependent (caller) classes and identify the target code to test.

We used 
\chan{PurityChecker-v3.1.2~\cite{purity-checker}} as pattern-based semantic-preservation checker, 
Randoop-v4.3.4~\cite{Randoop} as a test generation tool and  GPT-5~\cite{OpenAI_2025}, Deepseek-reasoner~\cite{deepseek} and LLama-3.3-70B-Instruct~\cite{llama} as generative large language models. 



All experiments were conducted on a regular machine equipped with an Intel Core i7-14700K processor (3.70 GHz, 6 cores, 2 threads per core), 96 GB of RAM, and an 8 TB HDD. The system was running 64-bit Ubuntu 22.04.5 LTS.

\section{Results}

\subsection{RQ1: Effectiveness of the \toolname}
\label{RQ1}

\begin{table}[t]
\vspace{-0.5em}
\centering
\caption{Average accuracy, precision and recall of \toolname (with ChatGPT). ExecFailures are considered as inaccurate predictions.}
\vspace{-0.5em}
\label{tab:RQ1_acc_prec_rec}
\begin{tabular}{ccccc}
\toprule
 & 
\textbf{Exec. Fail} & 
\textbf{Accuracy} & 
\textbf{Precision} & 
\textbf{Recall} \\
\midrule
\textbf{\toolname} & 
{3.82\%} &
{75.95\%} &
{100\%}  & 
{58.89\%}   \\
\bottomrule
\end{tabular}
\end{table}

To evaluate our approach, we run it on our dataset and compute its precision, recall and accuracy in distinguishing semantic-preserving from semantic-changing commits. We report the obtained results 
in Table~\ref{tab:RQ1_acc_prec_rec}.
As can be seen, 
\toolname executed correctly on the majority of the cases, except \chan{$\approx$3\%} of the commits on which either code- or test-generation failed.

More importantly, it can distinguish semantic-preserving from -changing commits in \chan{$\approx$76\%} of the cases, with a precision of \chan{100\%}. 
The \chan{100\%} scored precision indicates that all identified semantic-changing commits by \toolname are correctly classified and indeed semantic-changing.
The approach is however, unable to detect semantic changes in some commits, classifying them as preserving ones, and thus, scoring a recall of \chan{$\approx$59\%}. 
Out of the \chan{95} semantic-changing commits \toolname identified \chan{53} of them, while it achieved \chan{100\%} accuracy on the preserving ones. 
This difference is expected as writing tests to discover a behavioral change is typically more challenging than missing a non-existent one (case of semantic-preserving commits).

 \begin{tcolorbox}[colback=gray!10, colframe=gray!50,boxrule=0.4pt,arc=1pt,left=6pt,right=6pt,top=4pt,bottom=4pt,fontupper=\small,before skip=8pt,after skip=8pt]
\textbf{\toolname distinguishes accurately between semantic-preserving and semantic-changing commits in \chan{$\approx$76\%} of the cases, with a \chan{100\%} precision.}
\end{tcolorbox}
\vspace{1em}


\subsection{RQ2: Importance of LLM Selection}
\label{RQ2}





\begin{table}[!t]
\vspace{-0.5em}
\centering
\caption{\toolname variants relying on different LLMs}
\vspace{-0.5em}
\label{tab:RQ2_llms_variants}
\begin{tabular}{lcccccc}
\toprule
\textbf{Model} &
\textbf{Accuracy} &
\textbf{ExecFailure} &
\textbf{Precision} &
\textbf{Recall}&
\\
\midrule

{ChatGPT} &
{75.95\%} &
{3.82\%} &
{100\%}  & 
{58.89\%}  \\
\midrule

{DeepSeek}
&  63.93\% & 14.20\% & 100\% & 50.00\% \\
\midrule

{CodeLlama}
&  36.61\% & 36.61\% & 100\% & 19.67\%  \\
\bottomrule

\end{tabular}
\vspace{1em}
\end{table}

\begin{table}[t]
\vspace{-0.5em}
\centering
\caption{Number of commits for which the dependent code successfully compiled after each LLM call (try).\small{ When the LLM generates non compiling dependent-code, \toolname asks for a retry.}\ak{Maha: check number of commits(Done)}}
\label{tab:RQ1_iteration_llm}
\begin{tabular}{cccc}
\toprule
 & 
\textbf{Try 1} & 
\textbf{Try 2} & 
\textbf{Errors} \\
\midrule
{ChatGPT} & 
\text{164}  &
\text{11}  & 
\text{8}  \\
\midrule
\multirow{1}{*}{}
{DeepSeek} & 
\text{134}  &
\text{14}  & 
\text{35}  \\
\midrule
\multirow{1}{*}{}
{CodeLLama} & 
\text{80}  &
\text{23}  & 
\text{80}  \\
\bottomrule
\end{tabular}
\end{table}

Table~\ref{tab:RQ2_llms_variants} reports the execution results of \toolname variants; relying on different generative models, i.e., ChatGPT, DeepSeek and CodeLlaMA. 
Perhaps surprisingly, we observe large differences between the three variants performances. 
Particularly, we notice that the CodeLlaMa and DeepSeek variants fail to execute correctly in \chan{$\approx$14\% and 36\%} of the commits, respectively, which is about 4 and 9 times the \chan{$\approx$3\%} failure rate of the ChatGPT variant.
As can be seen in Table~\ref{tab:RQ1_iteration_llm}, both models fail often to generate compilable dependent-code even after two iterations. 
Consequently, these high execution failures caused the drop in the overall performance of these variants, achieving relatively lower accuracies of \chan{$\approx$36\% and $\approx$63\%} for CodeLlaMa and DeepSeek variants, respectively.

While the use of different LLMs in the code and test-generation phases had a big impact on the approach's accuracy, the impact is less noticeable on the precision. In fact, all variants kept a \chan{100\%} precision in semantic-changing commits identification and had lower precisions for the preserving ones, with ChatGPT the best performing model (\chan{$\approx69\%$}), followed closely by DeepSeek (\chan{$\approx65\%$}), then CodeLlaMa falling behind and scoring \chan{$\approx52\%$}.
Together with the large accuracy differences, these findings confirm that ChatGPT is an appropriate  model for our approach.

\ak{Maha we just add here the venn diragram and this way, we can show that ChatGPT is eventually subsuming the others, in a sense where any commits correctly identified by the others, is also identified by ChatGPT}

\begin{tcolorbox}[colback=gray!10, colframe=gray!50,boxrule=0.4pt,arc=1pt,left=6pt,right=6pt,top=4pt,bottom=4pt,fontupper=\small,before skip=8pt,after skip=8pt]
\textbf{ChatGPT outperforms DeepSeek and CodeLlaMa in code- and test-generation steps of \toolname, leading to the identification of \chan{4\%} and \chan{21\%} more semantic changing commits.}
\end{tcolorbox}

\subsection{RQ3: Importance of dependent code generation}
\label{RQ3}

  

To evaluate the contribution of the dependent code generation in exposing the behavioral difference between pre- and post-commit versions, we compare its performance in generating semantic-changes-revealing tests with that of the original developer-written dependent code. 

\begin{figure}[t]
\vspace{-1em}
    \centering
  \adjincludegraphics[width=0.7\linewidth, trim={{.08\width} {.05\width} {0.08\width} {.07\width}} ,clip]{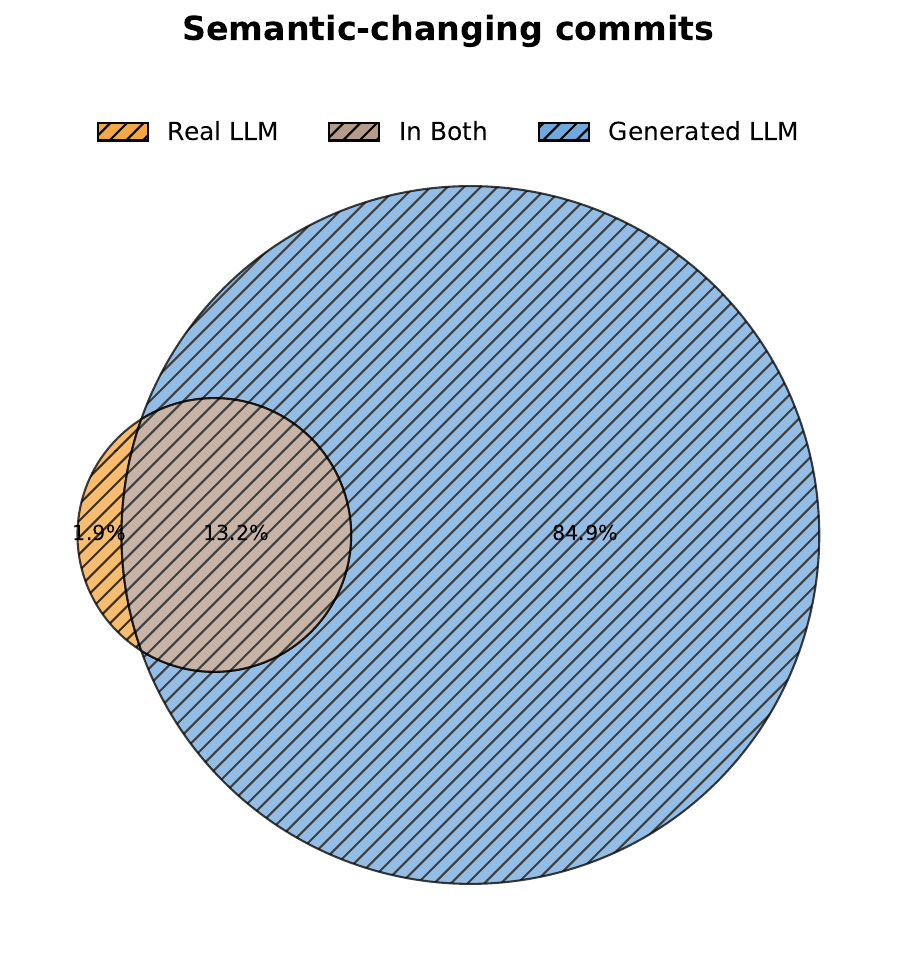}
  \vspace{-0.8em}
    \caption{Percentage of detected semantic changing commits when testing original and generated dependent classes.}\ak{Maha: put percentage instead of number of commits. divide by 52.(Done)}
    \label{fig:original-vs-generated-detected}
\vspace{-1em}
\end{figure}

We observe that tests targeting the generated code outperform 
those targeting original dependent code. 
From Figure~\ref{fig:original-vs-generated-detected} we can see that the generated code enables the identification of almost all semantic-changing commits, exposing over \chan{98\%} of them, while targeting the orignal dependent-code could only identify \chan{$\approx15$\%}.
Moreover, the majority of semantic-changing commits (\chan{$\approx$85\%}) can only be identified through the generated dependent code. 
This is interesting as it highlights that 
the generated code is responsible for the generation 
of semantic changes revealing tests, in the majority of the cases.


The analysis confirms that code generation is required for the semantic difference detection in \chan{50\%} of the cases, as the commit changed code is not called by any original code. 
It is noted that framework and utility projects, e.g. libraries, typically have not-calling code as it is intended to be used and called by host applications. 

\begin{figure}[t]
\vspace{-0.5em}
    \centering
  \adjincludegraphics[width=0.9\linewidth, trim={{.0\width} {.00\width} {0.00\width} {.00\width}} ,clip]{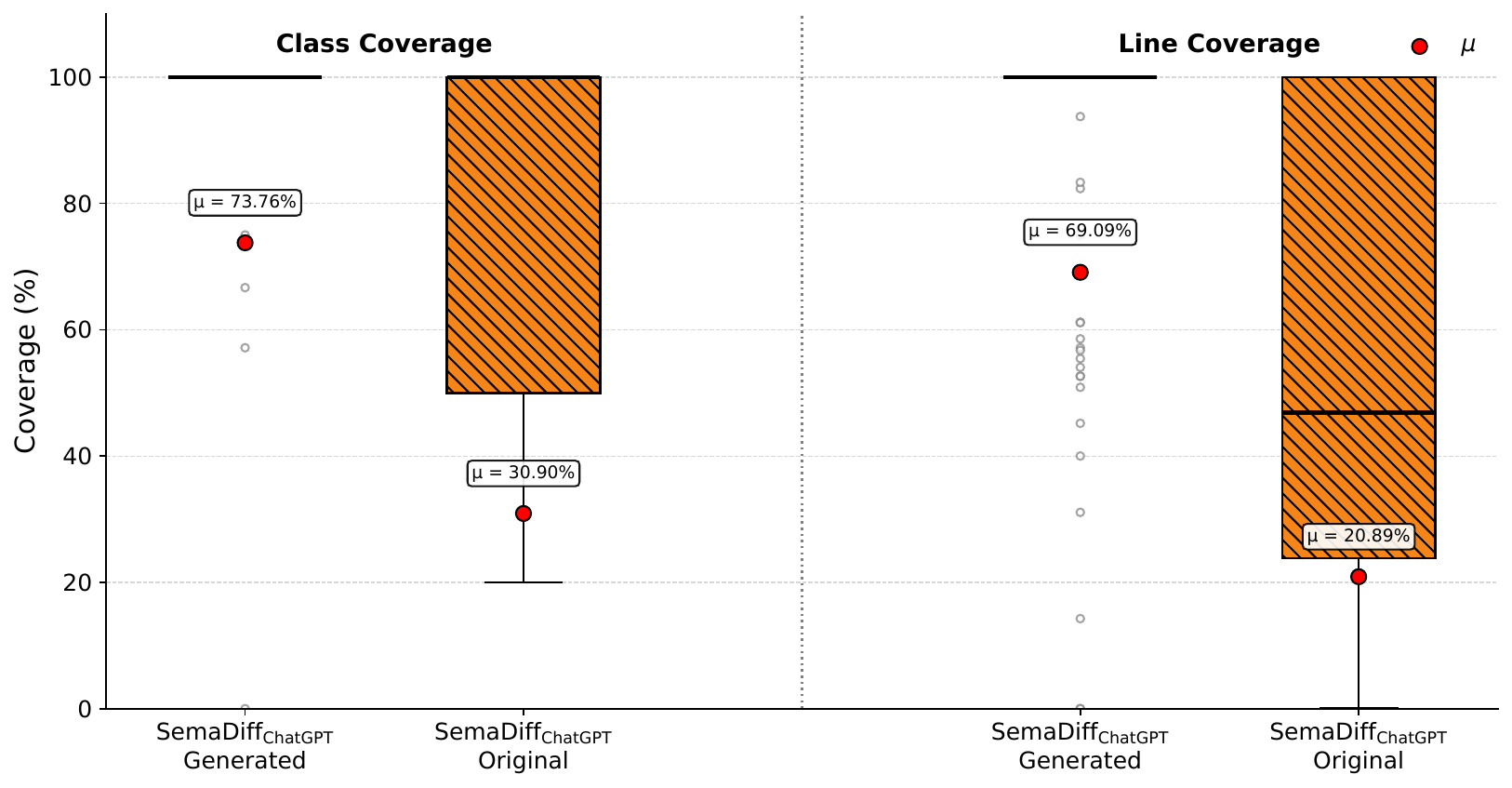}
    \caption{Code coverage of the modified code by test cases targeting the generated and original dependent code.} 
    \label{fig:chatgpt-lines-class-coverage}
\vspace{-1em}
\end{figure}

To have a better understanding of the difference between generated and original code performance in semantic-changes-revealing, we compute the generated-test code coverage. 
We notice that generated code ouperforms original one, leading to the generation of tests that execute on average about \chan{74\%} of dependent classes compared to about \chan{30\%} only by the original. 
\chan{Figure~\ref{fig:chatgpt-lines-class-coverage}} illustrates the code coverage of changed lines and classes by generated tests, when targeting each of the original and dependent code.
The alined boxes at 100\% confirm the large advantage of the generated code, achieving a noticeably higher coverage than that achieved by the original code, with respectively \chan{$\approx$42\%} and \chan{$\approx$48\%} more average class and line coverage.
Hence, the limited effectiveness of tests targeting original code in revealing semantic modifications can be explained by their lack of commit-change coverage.




By analyzing the code manually, we observe that, unlike the original classes, the generated ones are relatively simpler, including only methods that invoke the changed lines by the target commit, as requested by the LLM.
Moreover, the relevant calling function in the original code often appears deep within a complex execution path that depends on specific conditions that make hard its execution. For example, in Figure~\ref{fig:example_original_generated}, the extract variable refactoring is performed on the method \texttt{ProtocolReplyReceived} in which the written output is defined explicitly as a variable and later used in the method. In the real dependent class \texttt{IMAPExportMbox.java}, the method \texttt{protocolReplyReceived} is triggered only when:
(i) the listener is registered on the IMAP protocol client, and
(ii) a real network interaction occurs, and
(ii) replies are read from the socket during a sequence of IMAP commands. 
This means the function that invokes it might be inside a complex execution flow that includes network initialization, authentication, choosing the right folder, streaming the fetch, and handling errors. To reliably exercise the callback, tests require extensive setup (such as a real server, mocks, timing control, and precise protocol ordering), which in turn makes coverage difficult. In comparison to the generated dependent class that calls \texttt{protocolReplyReceived} directly, instead of running the full flow of the real dependent class.

 \begin{tcolorbox}[colback=gray!10, colframe=gray!50,boxrule=0.4pt,arc=1pt,left=6pt,right=6pt,top=4pt,bottom=4pt,fontupper=\small,before skip=8pt,after skip=8pt]
\textbf{Code generation plays an essential role in \toolname effectiveness. Generated code exposes best commit modifications, thereby forming an efficient target for the semantic-changes detection test generation and outperforming (subsuming) original dependent code; exposing \chan{$\approx98\%$} of identified semantic-changing commits among which \chan{$\approx84\%$} exclusively.    }
\end{tcolorbox}
\vspace{-0.5em}

\begin{figure}
\vspace{-1.3em}
    \centering
     \begin{subfigure}[b]{0.49\textwidth}
     \centering
\begin{lstlisting}[language=diff]
+ private static final String DIRECTION_MARKER_RECEIVE = "< ";
  public void protocolReplyReceived(final ProtocolReplyReceived(final ProtocolCommandEvent event){
    if (showDirection) {
-      writer.print("< ");
+      writer.print(DIRECTION_MARKER_RECEIVE);
    }

\end{lstlisting}
\vspace{-0.7em}
\caption{Example refactoring commit changes.}
\vspace{0.3em}
\label{subfig:example_ref_chg}
\end{subfigure}
     \begin{subfigure}[b]{0.49\textwidth}
     \centering
\begin{lstlisting}[firstnumber=1]

final PrintCommandListener listener = new PrintCommandListener(System.out, true) {
            // override
            public void protocolReplyReceived(final ProtocolCommandEvent event) {
                if (event.getReplyCode() != IMAPReply.PARTIAL) {
                    // This is dealt with by the chunk listener
                    @\textbf{super.protocolReplyReceived(}@                            @\textbf{event);}@
                }
            }
        };
\end{lstlisting}
\vspace{-0.7em}
\caption{Original code calling refactored code.}
\vspace{0.3em}
\label{subfig:example_real_code}
\end{subfigure}
     \begin{subfigure}[b]{0.49\textwidth}
     \centering
\begin{lstlisting}[firstnumber=1]
Public class Print {
    public String demoOutput(){
        StringWriter sw = new StringWriter();
        PrintWriter pw = new PrintWriter(sw);
        PrintCommandListener listener = new PrintCommandListener(pw, true, true);
           // Simulate receiving a reply
        ProtocolCommandEvent reply = new ProtocolCommandEvent(this, 230, "230 Welcome");
        @\textbf{listener.protocolReplyReceived(reply);}@
        return sw.toString()
}
\end{lstlisting}
\vspace{-0.7em}
\caption{Generated code by the LLM calling refactored code.}
\label{subfig:example_gen_code}
\end{subfigure}
\hfill
\vspace{-1em}
    \caption{Generated and developer-written code for calling a method changed by a refactoring commit. The generated code is a more suitable target for test generation as it exposes the changed-method calling instruction, unlike the developer code. 
    }
    \label{fig:example_original_generated}
    \vspace{-1em}
\end{figure}

\subsection{RQ4: Impact of test generation}

\ak{Maha: please update numbers and figures here (Done)}

To evaluate \toolname's LLM-based test generation component, we compare it with a well-established rule-based feedback-guided approach, i.e. Randoop, in terms of semantic-changing commits detection. 
To do so, we create a Randoop-based version of \toolname and run it 
five times on our dataset, 
targeting the same original and generated code for each commit. We report the results of the best run, achieving the highest overall accuracy (more details in Section~\ref{sec:setup}). 
\begin{table}[t]
\vspace{-1.2em}
\centering
\caption{Average accuracy, precision and recall of \toolnamerand (when generating tests with Randoop). ExecFailures are counted as inaccurate predictions.}
\label{tab:RQ4_rand_acc_prec_rec}
\begin{tabular}{ccccc}
\toprule
 & 
\textbf{Exec. Fail} & 
\textbf{Accuracy} & 
\textbf{Precision} & 
\textbf{Recall} \\
\midrule
\textbf{\toolnamerand} & 
\chan{45.35\%} &
\chan{34.9\%} & 
\chan{100\%}  & 
\chan{37.9\%}   \\
\bottomrule
\end{tabular}
\vspace{-1.5em}
\end{table}

\toolnamerand run successfully and generated tests for only for \chan{100} commits, failing in \chan{45.35\%} of the cases, as reported in Table~\ref{tab:RQ4_rand_acc_prec_rec}.
This $ExecFailure$ rate is almost \chan{12} times that of \toolname, contributing a lot in its accuracy drop to \chan{22.34\%}.  
We notice however that \toolnamerand scores a \chan{100\%} precision, same as \toolname when generating tests using generative language models. 
This is interesting and leads to the same conclusion obtained from answering RQ2~\ref{RQ2}, as it shows that even with less accurate test generators, \toolname keeps a \chan{100\%} precision, not predicting any semantic-preserving commit as changing.

\begin{figure}[t]
    \centering
  \adjincludegraphics[width=0.95\linewidth, trim={{.0\width} {.00\width} {0.00\width} {.00\width}} ,clip]{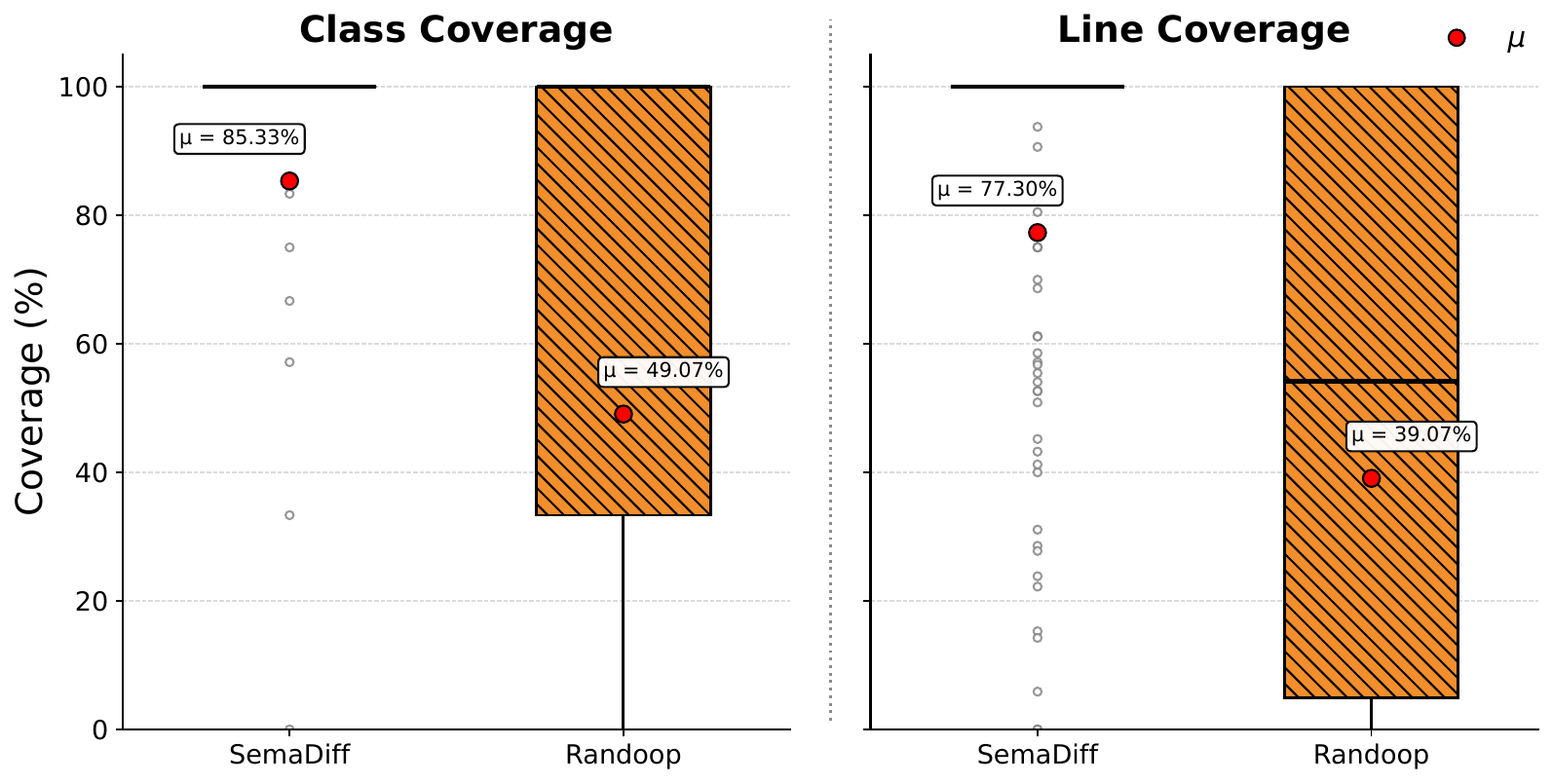}
\vspace{-0.45em}
    \caption{Average modified code coverage of tests generated by \toolname and Randoop, on the class- and line-level of granularity. \small{We include only commits for which the tool generated at least one test; \chan{100} commits for Randoop Vs \chan{176} for \toolname.}} 
    \label{fig:randoop-lines-class-coverage}
\end{figure}

In Figure~\ref{fig:randoop-lines-class-coverage}, we compare the modified-code coverage of tests generated by \toolname and Randoop, on the class- and line-level of granularity, including only commits for which every approach had run successfully (\chan{100} commits for Randoop Vs \chan{176} for \toolname).
The boxplots show that 
\toolname generates tests that reach the modified classes in median \chan{100\%} and average of over \chan{85\%} of commits which is largely higher than the \chan{49\%} average class coverage rates achieved by \toolnamerand.
We observe the same trend when analysing the modified lines coverage achieved by both approaches, 
with \toolnamerand achieving about half the coverage of \toolname. 
Hence, the large efficiency advantage of \toolname over \toolnamerand can be explained, not only by its lower $ExecFailure$ rates, but also by the higher coverage on the modified code. 

 Overall, \toolnamerand{} performed relatively poorly with an accuracy of \chan{$\approx22$\%}, running successfully in only \chan{$\approx$55\%} of the cases. 
 \chan{In addition, the analysis of joint and disjoint sets of accurately classified commits by both approaches, shows that \toolname subsumes \toolnamerand, classifying correctly all those accurately classified by \toolnamerand.} 
 These findings validate our approach design in generating tests using a generative model, instead of a traditional test-generation approach, i.e. Randoop. 

 \begin{tcolorbox}[colback=gray!10, colframe=gray!50,boxrule=0.4pt,arc=1pt,left=6pt,right=6pt,top=4pt,bottom=4pt,fontupper=\small,before skip=8pt,after skip=8pt]
\textbf{ChatGPT outperforms Randoop in generating tests with higher coverage of the modified code, enabling \toolname to achieve a much higher accuracy than \toolnamerand in distinguishing semantic-preserving from semantic-changing commits. The effectiveness variation of test generation component does not impact the approach precision, as both \toolnamerand and \toolname achieve a \chan{100\%} of precision.}
\end{tcolorbox}

\subsection{RQ5: Effectiveness of static-analysis approaches}
\label{RQ5}

We execute PurityChecker, and ask three LLMs to classify the commits from our dataset and report the computed results in Table~\ref{tab:purity_checker_accuracy_comparison}.
As can be seen, PurityChecker did not give any result in \chan{65.03\%} of the commits, and consequently distinguishing accurately between semantic-preserving and changing commits in only \chan{20.77\%} of the cases.
This can be explained by the fact that it supports only 9 types of refactoring and 
its approach is restricted to the commits syntactical changes (mapping changes with refactoring-type-patterns), without any behavioural exploration.
This explains also its low precision of 66.67\% compared to the 100\% achieved by \toolname. 
Nevertheless, PurityChecker remains an efficient lightweight approach, achieving high accuracy within its scope, when accounting only the commits for which it provided results.  

The three language models, not requiring to generate or execute any code or test, performed successfully on all the commits, achieving high accuracies, of \chan{78}, \chan{83} and \chan{86} for respectively CodeLlama, DeepSeek and ChatGPT. 
This is interesting as it shows that they are effecient at this task and can be used as a lightweigh solution to distinguish between semantic-changing and preserving commits.
However, although they achieve higher accuracies and recalls \chan{(between 76.84 and 91.58)} than \toolname, they achieve lower precision, ranging between \chan{81.11 and 84.47}.
This means that over \chan{15\%} of their predicted semantic-changing commits are actually preserving, and thus, limiting the reliability and usability as semantic-changing commits identifiers, compared to \toolname. 
Similarly, they mis-predict some semantic-preserving commits as changing, which is not the case of \toolname.
This confirms the complexity of the task and motivates further the aim of our approach -- identifying semantic-changing commits through dynamic analysis.


 \begin{tcolorbox}[colback=gray!10, colframe=gray!50,boxrule=0.4pt,arc=1pt,left=6pt,right=6pt,top=4pt,bottom=4pt,fontupper=\small,before skip=8pt,after skip=8pt]
\textbf{LLMs are efficient in distinguishing semantic-preserving commits from changing ones. However, their lack behavioral of analysis and semantic difference proof limits their detection certainty and consequently, not fulfilling \toolname's goal and not reaching its \chan{100\%} of precision.}
\end{tcolorbox}







\section{Threads to validity}

\begin{table}[htbp]
\centering
\caption{Prediction results of LLMs and PurityChecker.}
\vspace{-0.5em}
\label{tab:RQ5_llms_variants}
\begin{tabular}{lcccccc}
\toprule
\textbf{Approach} &
\textbf{Accuracy} &
\textbf{NoResult} &
\textbf{Precision } &
\textbf{Recall }\\
\midrule

{PurityChecker}

         & 20.77 & 65.03  & 66.67  & 76.19 \\
\midrule

{ChatGPT}

         & 86.89 & 0.00  & 84.47  & 91.58 \\
\midrule

{DeepSeek}

        & 83.61 & 0.00  & 82.83  & 86.32  \\
\midrule

{CodeLlama}
           & 78.14 & 0.00  & 81.11  & 76.84 \\
\bottomrule

\end{tabular}
\label{tab:purity_checker_accuracy_comparison}
\end{table}


Threats to external validity may arise from the use of external tools, such as large language models and test generation frameworks, in our validation study. To reduce this threat, we employ state-of-the-art tools, including three widely used generative language models, 
Randoop, a widely recognised tool for automated unit test generation, RefactoringMiner and its extension PurityChecker, implementing the state-of the art pattern-based refactoring analysis approaches. 
Moreover, we analyse every approach execution, considering their specific scopes, and reporting detailed corresponding results, e.g. accounting for their execution failures.


The implementation of \toolname{} relies on the Spoon library~\cite{PawlakMPNS16} for abstract syntax tree anaylsis and call graph construction. Therefore, we acknowledge that any potential bug or limitation in Spoon could impact \toolname functioning, e.g. reducing its effectiveness or the validity of our results. To mitigate this threat, we manually validated the correctness of multiple generated call graphs. 
Similarly, we tested the rest of \toolname components and the validation study pipeline.


A potential threat to the validity of this study concerns the data used for the evaluation, namely the selected commits. To mitigate this threat, we included a diverse set of recent commits with different refactoring types from widely used open-source projects. 
Furthermore, we manually inspected all commits, by more than one author, individually, then collectively in order to label each commit as semantically preserving or not. 
From the obtained results, particularly the different accuracies scored by all compared approaches and all below 90\%, we believe that the dataset contains various and challenging commits, not easily classified by all approaches.
We acknowledge however that our results could be different when targeting commits from other projects, written by other developers, or in other programming languages than Java.

\section{Conclusion}
We proposed \toolname, a novel approach that aims at identifying semantic changes in refactoring commits by generating dependent code along with corresponding tests and executing them. We evaluate our approach on a dataset of \chan{183} manually annotated commits containing both semantic-preserving and semantic-changing refactorings. The results demonstrate that our method achieves an average accuracy of \chan{$\approx$78\%} in validating whether refactoring commits preserve semantics or not. 
Although similar or less accurate than static-analysis LLM-based baselines (i.e. \chan{$\approx$78\%}, \chan{$\approx$84\%} and \chan{$\approx$87\%} for respectively CodeLlama, DeepSeek and ChatGPT), the approach outperforms them with a \chan{100\%} precision; always detecting correctly semantic changes.
Overall, the obtained results highlight the complexity of the task --distinguishing between semantic preserving and changing commits -- and endorses the use of \toolname to identify semantic changes.
\bibliographystyle{IEEEtran}
\bibliography{bibliography}

\end{document}